\documentclass[
showpacs,preprintnumbers,amsmath,amssymb,nofootinbib]{revtex4}
\usepackage{graphicx}
\usepackage{dcolumn}
\usepackage{bm}
\usepackage{amsmath,amssymb}
\usepackage{epsfig}
\def\d{\partial}
\newcommand{\be}{\begin{equation}}
\newcommand{\ee}{\end{equation}}
\newcommand{\bi}{\begin{itemize}}
\newcommand{\ei}{\end{itemize}}

\newcommand{\ba}{\begin{eqnarray}}
\newcommand{\ea}{\end{eqnarray}}
\newcommand{\gc}{\gamma_c}
\newcommand{\f}{\frac}

\def\g{\gamma}
\def\d{\partial}
\def\dd{\textrm{d}}

\def\abar{\bar\alpha}

\begin{document}

\bibliographystyle{h-physrev4}

\title{Universal behavior of the gluon saturation scale\\
at high energy including full NLL BFKL effects}

\author{Guillaume~Beuf}
\email{gbeuf@quark.phy.bnl.gov}
\affiliation{Department of Physics,\\
Brookhaven National Laboratory,\\
Upton, NY 11973, USA}

\begin{abstract}

The universal traveling wave solution to the Balitsky-Kovchegov equation with running coupling (and
other equations in the same universality class) is extended to subleading orders at large rapidity and small dipole size $r$.
The large rapidity expansion of the logarithm of the saturation scale $Q_s(Y)$ is derived from that traveling wave solution. In addition to
the two already known leading terms in $Y^{1/2}$ and $Y^{1/6}$, which are determined only by the LL BFKL
kernel, the three following terms in $Y^0$, $Y^{-1/6}$ and $Y^{-1/3}$ are obtained. They are universal and sensitive
to NLL BFKL effects. Initial condition dependence and NNLL BFKL effects would both start to appear
only at the next order, which is in $Y^{-1/2}$.
In the light of these results, the impact of the precise implementation of the running coupling in gluon saturation is discussed, \emph{i.e.} parent dipole size prescription \emph{vs.} Balitsky's prescription, and also the effect of the full NLL corrections, not yet implemented in numerical simulations.
\end{abstract}

\maketitle


\section{Introduction}
\label{sec:intro}

In the high energy limit for hadronic collisions, softer and softer partons are resolved in the hadronic wave-functions. According to the Balitsky-Fadin-Kuraev-Lipatov (BFKL) evolution \cite{Lipatov:1976zz,Kuraev:1977fs,Balitsky:1978ic}, these soft partons are emitted in a branching process \cite{Mueller:1993rr}. Eventually, the occupation number of the soft gluons becomes large, leading to the phenomenon of gluon saturation \cite{Gribov:1984tu,Mueller:1985wy,McLerran:1993ni,McLerran:1993ka,McLerran:1994vd} (for recent reviews see \emph{e.g.} \cite{Gelis:2010nm,Iancu:2003xm,Weigert:2005us,Balitsky:2001gj}). Further emission of softer gluons is a nonlinear collective effect, described by the Balitsky-Kovchegov (BK) \cite{Balitsky:1995ub,Kovchegov:1999yj,Kovchegov:1999ua} or JIMWLK\footnote{JIMWLK stands for Jalilian-Marian, Iancu, McLerran, Weigert, Leonidov and Kovner.} \cite{Jalilian-Marian:1997jx,Jalilian-Marian:1997gr,Jalilian-Marian:1997dw,Kovner:2000pt,Weigert:2000gi,Iancu:2000hn,Iancu:2001ad,Ferreiro:2001qy} equations, which generalize the BFKL evolution and overcome two of its drawbacks, \emph{i.e.} the violation of unitarity at fixed impact parameter and the diffusion into the infrared.
The boundary between the saturated part of a hadron wave-function, consisting of partons with small transverse momentum $k_\perp$ and/or large rapidity separation $Y$ and the dilute part is parameterized by the saturation scale $Q_s(Y)$.\\

Some properties of the solutions of the gluon saturation equations such as BK and JIMWLK can be derived analytically \cite{Gribov:1984tu,Iancu:2002tr,Mueller:2002zm,Munier:2003sj,Munier:2004xu}. The most powerful method for this purpose is borrowed \cite{Munier:2003vc} from the studies of nonlinear wave-front formation in some reaction-diffusion systems \cite{bramson,vanS}. The most interesting result is that many features of the solutions of the saturation equations are \emph{universal}, \emph{i.e.} independent of the initial condition, such as the large $Y$ behavior of the saturation scale, in agreement with numerical simulations \cite{GolecBiernat:2001if,Rummukainen:2003ns,Albacete:2004gw}.\\

Early phenomenological studies have found gluon saturation qualitatively consistent with the deep inelastic scattering
(DIS) data at low $x_{Bj}$ from HERA. For example, gluon saturation naturally explains the \emph{geometric scaling} property of the data \cite{Stasto:2000er}. Later, more obvious indications of gluon saturation have been found in particle production at forward rapidity in d-Au collisions at RHIC, such as the suppression of the Cronin peak  \cite{Arsene:2004ux,Adams:2006uz} or recently the disappearance of the recoil jet \cite{Braidot:2010zh} in two particles azimuthal correlations.  However, the phenomenological rapidity dependance of $Q_s(Y)$ extracted from the data is much slower than the one predicted from the BK and JIMWLK equations at leading logarithmic (LL) order.\\

That observation suggests to consider the effect higher order corrections to these saturation equations. Next-to-leading logarithmic (NLL) corrections to the BFKL equation \cite{Fadin:1998py,Ciafaloni:1998gs} are known and indeed large. The quark part of NLL corrections to the BK and JIMWLK equations has been obtained a few years ago \cite{Gardi:2006rp,Kovchegov:2006vj,Balitsky:2006wa}. Most of these corrections correspond to running coupling effects. One can then build improved LL BK and JIMWLK equations with running coupling at an appropriate scale \cite{Kovchegov:2006vj,Balitsky:2006wa}. Recently, the full NLL corrections to the BK equation have been derived \cite{Balitsky:2008zz,Balitsky:2009xg}.
State of the art phenomenological studies are now based on numerical simulations \cite{Albacete:2007yr} of the improved LL BK equation with Balitsky's running coupling prescription \cite{Balitsky:2006wa}, and provide a unified description
of the DIS data \cite{Albacete:2009fh}, single inclusive particle production \cite{Albacete:2010bs} and two-particles azimuthal correlations \cite{Albacete:2010pg} in d-Au collisions at RHIC, and two-particles long range rapidity correlations in Au-Au collisions at RHIC \cite{Dusling:2009ni}.\\

Analytical results are important to gain some insight into these numerical studies, in particular to understand the potential impact of the full NLL effects besides the running of the coupling, which are not yet implemented numerically. Previous studies \cite{Gribov:1984tu,Iancu:2002tr,Mueller:2002zm,Munier:2003sj} have shown that in the running coupling case, the solutions of the saturation equations seem still universal, but are in a different universality class than the fixed coupling ones, and the two leading terms in $\log Q_s(Y)$ at large rapidity have been obtained. It has been shown \cite{Peschanski:2006bm,Beuf:2007cw} that these two leading terms are independent of the details of the running coupling prescription and of the NLL (or higher order) contributions not associated with the running of the coupling. Hence, one has to calculate further subleading terms in the asymptotic behavior of the solutions and of $\log Q_s(Y)$ in order to discuss full NLL effects in a consistent way. That observation raises some questions about the validity of the previous attempt at obtaining NLL effects on the saturation scale evolution \cite{Triantafyllopoulos:2002nz}.\\

In this letter, the results for three new universal terms (\ref{coeff_c0},\ref{coeff_cm16},\ref{coeff_cm13}) in the asymptotic expansion \eqref{result_Ls_RC} of $\log Q_s(Y)$ in the running coupling case are presented in section \ref{sec:formResRC} and their derivation is outlined. These new terms are NNLL independent, but two of them are NLL dependant. The corresponding asymptotic expression for the saturation scale is used in section \ref{sec:plots} to compare the outcome of various saturation equations: the LL BK equation with the parent dipole size prescription or with Balitsky's prescription for the running coupling, and the NLL BFKL equation in momentum space supplemented by saturation effects. But first, let us start in the next section by a review of the known results in the fixed coupling case, as a warm-up.

\section{Review of the fixed coupling case}
\label{sec:FC}

At leading logarithmic (LL) order, the gluon saturation equations such as BK or JIMWLK can be written formally as\footnote{For simplicity, let us assume here that the target is homogeneous is the transverse plane. However, the traveling-wave method discussed here is valid in the general case of scattering at non-zero transfer on an inhomogeneous target\cite{Marquet:2005qu,Marquet:2005zf}. In this paper, the gluon saturation equations including gluon number fluctuations are not considered. These fluctuations occurring in the dilute regime have a big impact at fixed coupling \cite{Iancu:2004es}. When the coupling is running, these fluctuations should have similar effects in the high rapidity limit \cite{Dumitru:2007ew,Beuf:2007qa}, but are completely negligible in the relevant rapidity range.}
\begin{eqnarray}
\d_Y {N}(L,Y) &=& \abar\, \Big\{ \chi(-\d_{L}) {N}(L,Y) -  \,  \textrm{Nonlinear damping}\Big\}\label{BKhomkFC}
\end{eqnarray}
with the notation $\abar = \alpha_s N_c /\pi$, both in position space or momentum space. In the former case, $N(L,Y)$ is the dipole-target amplitude for a dipole of size $r$ and a rapidity interval $Y$, with $L=-\log (r^2 {Q_0}^2/4)$ and $Q_0$ an arbitrary reference scale. In the latter case, $N(L,Y)$ is typically an unintegrated distribution of gluons with transverse momentum $k_\perp$ and rapidity $Y$ in the target, with $L=\log ({k_\perp}^2/{Q_0}^2)$. The linear part of equation \eqref{BKhomkFC} involves the BFKL kernel at LL accuracy and restricted to zero conformal spin, whose characteristic function is
\be
\chi(\g)=2 \Psi(1)-\Psi(\g)-\Psi(1\!-\!\g)\label{chiLL_BFKL}
\ee
in both the position and momentum spaces, $\Psi(\g)$ being the digamma function.\\

The BFKL linear term in \eqref{BKhomkFC} implies an exponential instability in $Y$ of the vacuum solution ${N}(L,Y)=0$ and a diffusive behavior in $L$-space. The non-linear damping term in equation \eqref{BKhomkFC} tames the instability when $N(L,Y)$ becomes of order $1$. More precisely, we have $N(L,Y)\leq 1$ for the dipole-target amplitude in position space. For its Fourier transform as defined in Ref.\cite{Kovchegov:1999ua}, 
the exponential growth becomes a linear growth in $Y$ in the deep saturation regime. By contrast, the effective unintegrated gluon distribution appearing in the $k_\perp$-factorization for gluon production in DIS and pA turns over in the nonlinear regime and decreases back towards zero \cite{Kharzeev:2003wz}. For any of these cases, one expects a perturbative power tail in $k_\perp$ or $r$ in the UV, meaning a large-$L$ exponential tail of $N(L,Y)$. Hence the nonlinear regime of saturation is reached in the IR at lower values of $L$ first. At large enough finite values of $Y$, the nonlinear term in equation \eqref{BKhomkFC} is thus negligible for large $L$ but not for small $L$. The typical boundary value between these two regimes $L_s(Y)\equiv\log({Q_s(Y)}^2/{Q_0}^2)$ defines the saturation scale $Q_s(Y)$. Due to the exponential rise of $N(L,Y)$ with $Y$ at large $L$, $Q_s(Y)$ is expected to grow with the rapidity $Y$. Hence, the relevant solution for $N(L,Y)$ is a traveling wave front in $L$-space, where $\abar Y$ plays the role of the \emph{time} variable. At the time $\abar Y$, the wave front is at the position $L=L_s(Y)$ and moves towards larger $L$.\\

As noticed in \cite{Munier:2003vc}, the properties of exponential instability and diffusion in the linear regime and of non-linear damping are shared by both the equation \eqref{BKhomkFC} and the Fisher and Kolmogorov-Petrovky-Piskounov (FKPP) equation \cite{fish,KPP}, which can actually be written as \eqref{BKhomkFC} with $\chi(\g)=1+\g^2$ and a quadratic or a cubic nonlinear damping. Hence some mathematical results \cite{bramson,vanS} about nonlinear wave front formation for the FKPP equation remain valid in general for equation \eqref{BKhomkFC}.
 Both of these equations admit a special family of exact scaling solutions $N(L,Y)=N_\g (s_v(L,Y))$, also called uniformly translating front solutions, with
\ba
\textrm{the scaling variable}& & s_v(L,Y)=L- v \abar Y\label{varScFamilyFC}\, ,\\
\textrm{a scaling function with a tail}& & N_\g (s_v) \propto e^{-\g s_v} \quad \textrm{for large positive } s_v\label{TailFamilyFC}\, ,\\
\textrm{and the dispersion relation}& & v(\g)=\f{\chi(\g)}{\g}\label{RelDisp}\, .
\ea
An important feature is that the wave-front velocity $v(\g)$ admits a positive global minimum $v_c=v(\gc)$ in the relevant range for $\g$, which is $0<\g$ in the FKPP case and $0<\g<1$ when using the BFKL eigenvalue \eqref{chiLL_BFKL}. The scaling solution $N_{\gc} (s_{v_c}(L,Y))$ with minimal velocity is called the critical solution. The minimization condition for the velocity
\be
\chi(\gc)=\gc\, \chi'(\gc)\label{Eq_gc}
\ee
gives $\gc=1$ and $v_c=2$ in the FKPP case and $\gc\simeq 0.6275$ and $v_c\simeq 4.883$ in the QCD case.\\

The generic solutions of the FKPP and BK equations, outside that family of uniformly translating fronts, fall into two classes \cite{bramson,vanS}.
\bi
\item If the initial condition has a flatter tail than the critical solution at large $L$, then the long $Y$ behavior of $N(L,Y)$ is very sensitive to the initial condition.
\item If the initial condition has a steeper tail than the critical solution, $N(L,Y)$ converge to the critical front solution at large $Y$, and is called a \emph{pulled} front solution. Moreover, the convergence happens first in the fully nonlinear regime ${N}(L,Y)={\cal O}(1)$, and propagates in an universal way into the tail via diffusion. Hence, the dynamics of the solution is determined by the existence of the nonlinear term in equation \eqref{BKhomkFC} even in a wide kinematical region where its magnitude is actually negligible compared to the magnitude of the linear terms.
\ei
In QCD, the tail of the initial condition should be of the type $\exp(-L)$, due to the property of color transparency, which is steeper than the critical tail $\exp(-\gc L)$. The physically relevant solutions for QCD are thus of the pulled front type.\\

In Ref.\cite{vanS}, a method has been developed in order to study universal properties of the pulled front solutions of equations in the universality class of the FKPP equation and of equation \eqref{BKhomkFC}.
Large $Y$ asymptotic expansions of the pulled front solutions can be constructed in two different regimes. The first one is the \emph{front interior} regime where $L\!-\!L_s(Y)$ is kept fixed when $Y\rightarrow\infty$, corresponding to describe the wave-front in its rest frame. The second one is the \emph{leading edge} regime where $L\!-\!L_s(Y)={\cal O}(\sqrt{\abar Y})$ when $Y\rightarrow\infty$, corresponding to the crossover between the universal region of the front and the part of the tail still driven by the initial condition. Matching the two expansions together and imposing some causality requirement in the leading edge regime allow to determine most of the parameters and extract universal information \cite{vanS}. For example, $N(L,Y)$ depends only on the scaling variable $L\!-\!L_s(Y)$ in the increasing window $L\!-\!L_s(Y)\ll \sqrt{2\, \chi''(\gc)\, \abar Y}$. Combined with the appropriate dipole or $k_\perp$-factorization formula, gluon saturation thus explains the geometric scaling property seen in the low-$x$ HERA data \cite{Stasto:2000er}. One also obtain the universal asymptotic expansion of the position of the front \cite{vanS}, related to the saturation scale,
\be
L_s(Y)\equiv\log\left(\f{{Q_s(Y)}^2}{{Q_0}^2}\right)=v_c\, \abar Y -\f{3}{2 \gc} \log(\abar Y)+ \textrm{Const.}-\f{3}{\gc^2}\sqrt{\f{2 \pi}{\chi''(\gc)\,  \abar Y}} + {\cal O}\left(\f{1}{\abar Y}\right)\label{result_Ls_FC}\, .
\ee\\

Within QCD context, the first term in the expansion \eqref{result_Ls_FC} was derived in \cite{Gribov:1984tu,Iancu:2002tr}, the second one in \cite{Mueller:2002zm,Munier:2003sj} and last universal term in \cite{Munier:2004xu}. Due to the $L$-translational invariance of equation \eqref{BKhomkFC}, traducing the conformal invariance of high energy QCD at LL accuracy, the constant term in \eqref{result_Ls_FC} is directly sensitive to the initial condition, instead of being universal. Moreover, equation \eqref{BKhomkFC} is also invariant under $Y$-translations, and such translations modify the large $Y$ expansion of $L_s(Y)$, starting at the order ${\cal O}(\abar Y)^{-1}$. Hence, in the expansion \eqref{result_Ls_FC}, the coefficients of the terms of order ${\cal O}((\abar Y)^{-1})$ or further cannot be universal.\\

In Ref.\cite{Enberg:2006aq}, the effect of NLL corrections to the kernel of equation \eqref{BKhomkFC} has been studied, keeping the coupling fixed. It amounts to deform the dispersion relation \eqref{RelDisp}, and thus to shift the critical parameters $\gc$ and $v_c$ by corrections proportional to $\abar$. Hence, all the coefficients in the expansion \eqref{result_Ls_FC} are shifted each time one goes to the next logarithmic order in perturbation theory, LL, NLL, NNLL, etc. That property is true for gluon saturation in N=4 SYM theory. However, as we will see in the rest of this paper, this is not the case in full QCD, where running coupling effects stabilize the asymptotic expansion of $\log Q_s$ with respect to other higher order corrections.

\section{Saturation equations with running coupling}

The most important correction in QCD to the gluon saturation equations of the type \eqref{BKhomkFC} is the running of the coupling $\abar$. That effect corresponds an all-order resummation of a tower of terms in the perturbative expansion. However, as a first step, it is customary to simply promote the coupling $\abar$ in equation \eqref{BKhomkFC} to the one-loop expression
\be
\abar \mapsto \f{1}{b L}\, , \quad \textrm{where} \quad b=\f{11 N_c\!-\!2 N_f }{12 N_c}\label{OneLoopRunC}\, ,
\ee
without trying to calculate this tower of perturbative corrections. In momentum space, that prescription corresponds to take the running coupling at the scale of the parent gluon $k_\perp$. By contrast, when applying the prescription \eqref{OneLoopRunC} to an equation \eqref{BKhomkFC} in position space, such as the BK equation, one obtains the running coupling at a scale related to the parent dipole size $r$. By consistency, in the running coupling case one should take  the reference scale $Q_0$ to be $\Lambda_{QCD}$ in the appropriate renormalization scheme, so that now
\be
L=\log \left(\f{{k_\perp}^2}{{\Lambda_{QCD}}^2}\right)   \quad \textrm{or} \quad   L=-\log \left(\f{r^2 {\Lambda_{QCD}}^2}{4}\right)\, .\label{L_RC}
\ee
The gluon saturation equations with this type of running coupling prescription have been intensively studied numerically, and are suitable for our analytical calculations.\\

It is possible to write in a formal way the all order generalization of the gluon saturation equations \eqref{BKhomkFC}, following this prescription, as an asymptotic series in $(b L)^{-n}$
\begin{eqnarray}
\d_Y {N}(L,Y) &=& \frac{1}{b L}\, \Big\{ \chi(-\d_{L}) {N}(L,Y) -  \,  \textrm{Nonlinear damping}\Big\}\nonumber\\
& & +\frac{1}{(b L)^2}\, \Big\{\chi_{NLL}(-\d_{L}) {N}(L,Y) -  \,  \textrm{Nonlinear damping}\Big\}\nonumber\\
& & + \dots\, .\label{BKhomkRUN}
\end{eqnarray}\\

The usual form of the improved LL BK equation with Balitsky's running coupling prescription \cite{Balitsky:2006wa} is difficult to handle directly with our analytical methods. However, it is possible to expand the running couplings in order to rewrite that equation as a series of the type \eqref{BKhomkRUN} \cite{Beuf:2007cw}, but this time with the NLL contribution
\be
\chi_{NLL}(\g)=\f{b}{2} \left[\chi(\g)^2\!-\!\chi'(\g) \!-\! \f{4}{\g}\, \chi(\g)\right]\label{chiNLL_Bal}\, .
\ee
Since only part of the NLL corrections are included in that case, the renormalization scheme is not really fixed. However, it is natural to rescale $\Lambda_{QCD}$ with respect its $\overline{MS}$ scheme value as
\be
\Lambda_{QCD} =e^{-\Psi(1)}\, \Lambda_{QCD}^{\overline{MS}}\, .\label{LambdaQCDpos}
\ee
This choice corresponds to the position space analog of the usual $\overline{MS}$ scheme.\\

Due to some specific technical issues, the full NLL BK equation \cite{Balitsky:2008zz} will not be considered here, and its study is postponed to further works. Let us consider instead the NLL BFKL equation \cite{Fadin:1998py} in momentum space for the unintegrated gluon distribution, and add gluon saturation effects to obtain an equation of the type \eqref{BKhomkRUN}. In that case, the NLL eigenvalue $\chi_{NLL}(\g)$ writes\footnote{Notice that our notations differ from the ones of Ref.\cite{Fadin:1998py} by exchange of $\g$ and $1\!-\!\g$.}
\ba
\chi_{NLL}(\g)&=& \f{1}{2}\, \chi(\g) \chi'(\g)-\f{1}{4} \chi''(\g) +\f{3}{2}\, \zeta(3)-\phi(\g)  -\f{b}{2} \left[\chi(\g)^2\!-\! \chi'(\g) 
\right]+\left[\f{5b+1}{3}\!-\!\f{\pi^2}{12}\right]\chi(\g) 
\nonumber\\
& &-\f{\pi^2 \cos (\pi \g)}{4 (1\!-\!2\g)\sin^2 (\pi \g)}\left[3\!+\!\left(1\!+\!\f{N_f}{N_c^3}\right)\f{(2\!+\!3\g (1\!-\!\g))}{(3\!-\!2\g)(1\!+\!2\g)}\right]+\f{\pi^3}{4 \sin (\pi\g)}\label{chiNLL_FaLi}\, ,
\ea
in the $\overline{MS}$ scheme and defining the rapidity as $Y=\log(s/{k_\perp}^2)$, with the notation
\be
\phi(\g)=\sum_{n=0}^\infty (-1)^n \left[\f{\Psi(n\!+\!1\!+\!\g)\!-\!\Psi(1)}{(n\!+\!\g)^2}+\f{\Psi(n\!+\!2\!-\!\g)\!-\!\Psi(1)}{(n\!+\!1\!-\!\g)^2}  \right]\, .
\ee\\


\section{The saturation scale for traveling waves with running coupling}\label{sec:formResRC}

The mechanism of universal pulled front formation from steep enough initial conditions has been established
rigorously \cite{bramson} in the case of the FKPP equation, and remains valid for equations within the same universality class, such as the BK equation at LL accuracy. By contrast, the extension to the running coupling case is far from obvious. There is an evidence from numerical simulations \cite{Rummukainen:2003ns,Albacete:2004gw} that some analogous universal wave front formation holds, but no mathematical proof.\\

Qualitatively, one can justify such universality. Most of the ingredients allowing for pulled front solutions of the FKPP equation are still present in equation \eqref{BKhomkRUN}, such as the instability of the vacuum, the diffusion effects, the nonlinear damping, and also the steepness of the initial conditions relevant for QCD.
\emph{A priori}, the only missing ingredient is the family of uniformly translated front solutions $N(L,Y)=N_\g (s_v(L,Y))$ with (\ref{varScFamilyFC},\ref{TailFamilyFC},\ref{RelDisp}). Indeed, the saturation equation with running coupling \eqref{BKhomkRUN} does not admit any exact scaling solution. In the literature \cite{Mueller:2002zm,Munier:2003sj}, a family of approximate scaling variables of the type $s_v(L,Y)=L\!-\!f_v(Y)$ is usually considered, and then the methods developed in the fixed coupling case are applied in the running coupling case from this starting point. However, such choice of approximate scaling variable is equivalent, in some sense, to impose by hand the geometric scaling of the solution, and moreover it is not unique \cite{Beuf:2008mb}. It is thus essential to consider also other types of approximate scaling variables for equation \eqref{BKhomkRUN}. On the one hand, any dependance of the final results, such as the expression for $Q_s(Y)$, on the choice of approximate scaling variable would refute the universality property in the running coupling case. Indeed, the physical $Q_s(Y)$ should not depend on arbitrary choices made when writing the asymptotic expansions of the solution $N(L,Y)$. On the other hand, the geometric scaling property is often argued to be due to gluon saturation, and indeed arises dynamically from gluon saturation in the fixed coupling case, as discussed in section \ref{sec:FC}. It is thus important to understand if and how such a scaling property arises dynamically in the running coupling case at least approximately, without imposing it by hand.\\

Generically, an approximate scaling variable for the equation \eqref{BKhomkRUN} is given by  any function $s_v(L,Y)$ whose partial derivatives write \cite{Beuf:2008mb}
\be
\f{\d s_v}{\d L}=1+\dots \qquad \textrm{and} \qquad \f{\d s_v}{\d Y}= -\abar(L)\, v \left[1+\dots\right] = -\f{v}{b L}\left[1+\dots\right]\, ,\label{ScalingConditionRC}
\ee
where the dots stand for terms which go to zero when $L,Y\rightarrow\infty$ with $s_v(L,Y)/L \rightarrow 0$. For each value of $v$, there is an infinity of such approximate scaling variables $s_v(L,Y)$. The dispersion relation \eqref{RelDisp} is found to hold independently of the corrective terms in \eqref{ScalingConditionRC}. Hence, in the running coupling case there is a critical solution, which approximately scales with any variable $s_{v_c}(L,Y)$ of the type \eqref{ScalingConditionRC} with the same critical parameters $\gc$ and $v_c$ as in the fixed coupling case.\\

By analogy with the fixed coupling case, the solutions with steep enough initial conditions should converge to that
critical solution. And they should approximately scale with a variable $s(L,Y)$ given by $s_{v_c}(L,Y)$ supplemented by subleading diffusive terms encoding the relaxation towards the critical solution. At that point, it is possible to follow the method of Ref.\cite{vanS}. One can construct the front interior and the leading edge asymptotic expansions of those solutions in the dilute domain. In both of these expansions, $L^{-1/3}$ can be taken as the small parameter, and the trivial $\exp(-\gc\, s)$ contribution is factored out of the solution. The front interior expansion is done at fixed value of $s(L,Y)$, and the coefficients are polynomials in $s(L,Y)$. The leading edge expansion, by contrast, is done at fixed value of $z\propto s(L,Y)/L^{1/3}$, and the coefficients are more complicated functions of $z$, involving the Airy function $\textrm{Ai}(z)$ or its derivative, times polynomials. One rejects the possibility of having the exponentially growing Airy function $\textrm{Bi}(z)$ appearing in the leading edge expansion in order to allow for a smooth connection with the non-universal tail of the solution, by analogy with the FKPP case \cite{vanS}. From this requirement and the matching of the two asymptotic expansions, one determines most of the parameters appearing in these expansions and in the approximate scaling variable $s(L,Y)$. The details of this calculation as well as the asymptotic expressions obtained for the solutions $N(L,Y)$ will be reported and discussed elsewhere \cite{BeufToAppear}.\\

The saturation scale $Q_s(Y)$ and its logarithm
\be
L_s(Y) \equiv \log \f{Q_s^2(Y)}{\Lambda_{QCD}^2}\label{log_Qs}\, ,
\ee
are defined by a level line
\be
N(L_s(Y),Y)=\kappa\label{LevelLineQs}
\ee
where $\kappa$ is a given small constant \emph{e.g.} $\kappa=0.1$ or $\kappa=0.01$.
Using the front interior expansion constructed for $N(L,Y)$, one can solve the equation \eqref{LevelLineQs} explicitly, and obtain the universal large-$Y$ asymptotic expansion
\be
L_s(Y)=\left(\f{2 v_c Y}{b}\right)^{1/2}+\f{3 \xi_1}{4} \left(D^2\, \f{2 v_c Y}{b}\right)^{1/6}+c_{0}+c_{-1/6} \, \left(D^2\, \f{2 v_c Y}{b}\right)^{-1/6}+c_{-1/3} \, \left(D^2\, \f{2 v_c Y}{b}\right)^{-1/3}+ {\cal O}\left(Y^{-1/2}\right)\label{result_Ls_RC}\, .
\ee
The first term in \eqref{result_Ls_RC} was derived in \cite{Gribov:1984tu,Iancu:2002tr} and the second one in \cite{Mueller:2002zm,Munier:2003sj}, whereas the three next ones constitute the main result of the present study:
\ba
c_{0}&=& \Sigma-\f{1}{\gc}-\f{K_3}{2}+N_0\label{coeff_c0}\\
c_{-1/6}&=&{\xi_1}^2\, \left[\f{1}{72\, \gc^2}-\f{D}{32}+\f{K_3}{4 \gc}-\f{3}{8}\, {K_3}^2+\f{3}{10}\, K_4\right]\label{coeff_cm16}\\
c_{-1/3}&=&\xi_1 \Bigg\{\f{\Sigma^2}{6 \gc}+\left[\f{K_3}{2\gc}\!-\!\f{1}{3\gc^2}\right]\Sigma +\f{2}{27\gc^3}+\f{D}{3\gc}+K_3\, \left[-\f{2}{3\gc^2}\!+\!\f{3 K_3}{4\gc}\!-\!\f{11 {K_3}^2}{4}\right]\nonumber\\
& &\qquad +K_4\, \left[-\f{1}{\gc}\!+\!6 K_3\right]-3 K_5  + \f{1}{2\gc^2}\left[1\!+\!3\gc K_3\right]\left[N_0\!-\!\gc N_1\right]+N_2\!-\!D N_0\Bigg\}\label{coeff_cm13}\, .
\ea
In the previous relations $\xi_1$ stands for the rightmost zero of the Airy function, \emph{i.e.} $\xi_1\simeq -2.338$, and the following notations have been used:
\be
D\equiv \f{\chi''(\gc)}{2\, \chi(\gc)}=\f{\chi''(\gc)}{2\, \gc v_c}\, , \qquad
K_n\equiv \f{2\, \chi^{(n)}(\gc)}{n! \, \chi''(\gc)}\, \quad \textrm{for}\quad n\geq 3\, , \qquad \textrm{and} \qquad
N_n\equiv \f{\chi_{NLL}^{(n)}(\gc)}{n!\, \chi(\gc)\, b}\, \quad \textrm{for}\quad n\geq 0\, .\label{D_Kn_Nn}
\ee
The parameter $D$ measures the diffusion rate with respect to the instability growth. Each parameter $K_n$ contains the $n$th derivative of the LL eigenvalue $\chi(\g)$, so that the $K_n$'s correspond to the contributions beyond the diffusive approximation of the LL kernel which is sometimes performed. The parameters $N_n$ quantify the contributions of the NLL eigenvalue $\chi_{NLL}(\g)$ or its derivatives to the expansion \eqref{result_Ls_RC}.\\

In the expressions \eqref{coeff_c0} and \eqref{coeff_cm13}, $\Sigma$ is a parameter depending on the height $\kappa$ at which the saturation scale is defined via \eqref{LevelLineQs}. It can be expressed as
\be
\Sigma=-\f{1}{\gc}\, \textrm{W}_{-1}\left(-\f{\gc\, \kappa}{{\cal A}\, e}\right)\, ,\label{sol_Sigma}
\ee
where $\textrm{W}_{-1}(x)$ is the $-1$ branch of the multivalued Lambert function $W(x)$ \cite{LambertFunc}, and ${\cal A}$ is the overall normalization constant of $N(L,Y)$ appearing in the front interior and the leading edge expansions. The calculations of these two expansions are performed using the linearized version of the equation \eqref{BKhomkRUN}, so that ${\cal A}$ is undetermined at that stage. The expression of $N(L,Y)$ obtained from the front interior expansion has a maximum and decreases in the IR. This maximum signals the breakdown of the front interior expansion due to the onset of the fully nonlinear regime. The normalization constant ${\cal A}$ should then be determined by matching numerically the front interior expansion to full numerical solutions extending into the deep saturation regime. Hence, ${\cal A}$ is determined by the height of $N(L,Y)$ at which the nonlinear damping terms in the equation \eqref{BKhomkRUN} become relevant. The constant ${\cal A}$ is thus independent of the initial condition, as long as the latter is steep enough to have convergence towards the critical solution, and is also independent of the presence of NLL or higher order terms in the equation, apart from the running of the coupling.
\emph{A priori}, one can expect a value of order one for ${\cal A}$. However, this value should depend on the considered quantity: dipole target amplitude, effective or physical unintegrated gluon distribution, which have a different behavior in the deep saturation regime. The formula \eqref{sol_Sigma} is valid only for $\gc\, \kappa\ll {\cal A}$, \emph{i.e.} when the saturation scale is defined on a level line \eqref{LevelLineQs} far enough from the deep saturation regime.\\

Due to many cancelations occurring either when constructing the front interior and leading edge expansions, or when extracting $L_s(Y)$ from them, the result \eqref{result_Ls_RC} does not depend on the choice of approximate scaling variable $s(L,Y)$ used to write the asymptotic expansions of $N(L,Y)$. As discussed previously, this is a non-trivial consistency check of our method. Hence, we have a new strong hint that in the running coupling case also, the solutions with steep enough initial condition converge universally towards a critical one, despite the absence of a family of uniformly translating front solutions.\\

Translations in $Y$ would generate contributions starting at order $Y^{-1/2}$ in the expansion \eqref{result_Ls_RC} out of the first term. Due to this remark and to the $Y$-translational invariance of equation \eqref{BKhomkRUN}, the coefficients of the expansion \eqref{result_Ls_RC} cease to be universal at order $Y^{-1/2}$. By contrast, translations in $L$ are no longer a symmetry of the equation \eqref{BKhomkRUN} due to the running of the coupling, \emph{i.e.} due to the conformal anomaly. Hence, all the universal terms have been derived in the expansion \eqref{result_Ls_RC}. It has been anticipated in Ref.\cite{Mueller:2003bz} that, in the running coupling case, the constant term $c_{0}$ should be independent of the initial condition, and that initial condition effects should be delayed until the order $Y^{-1/2}$. This prediction, which was already found consistent with numerical simulations \cite{Rummukainen:2003ns,Albacete:2004gw}, is thus confirmed by direct calculation. In practice, that property implies that the nuclear enhancement of the saturation scale $Q_s$ by a factor $A^{1/6}$ is progressively washed out by the rapidity evolution, so that at very high energy all hadrons and nuclei have the same saturation scale, with a normalization set by the conformal anomaly of QCD and not by the nature of the target. By contrast, the nuclear enhancement of $Q_s$ would survive at high energy in a conformal Yang-Mills theory, hidden in the non-universal constant term of the expansion \eqref{result_Ls_FC}.\\

The front interior and the leading edge expansions are asymptotic expansions in powers of $L^{-1/3}$. It is thus clear that, apart from the running of the coupling, NLL BFKL effects may start appearing in the expansion \eqref{result_Ls_RC} of $L_s(Y)$ at order $Y^0$, NNLL BFKL effects at order $Y^{-1/2}$, NNNLL BFKL effects at order $Y^{-1}$, and so on \cite{Peschanski:2006bm,Beuf:2007cw}. The universal terms can thus depend only on the known LL and NLL BFKL kernels. On the contrary, the fact that the coefficient $c_{-1/6}$ \eqref{coeff_cm16} is independent of NLL contributions is an unexpected outcome of our calculation.\\

The coefficient $c_{-1/3}$ \eqref{coeff_cm13} depends on NLL BFKL effects only through the combinations $N_0\!-\!\gc N_1$ and $N_2\!-\!D N_0$. It is instructive to notice that, if the NLL characteristic function $\chi_{NLL}(\g)$ is proportional to the LL one $\chi(\g)$, both $N_0\!-\!\gc N_1$ and $N_2\!-\!D N_0$ vanish. Different choices of renormalization scheme give characteristic functions $\chi_{NLL}(\g)$ differing by terms proportional to $\chi(\g)$. Hence, the coefficient $c_{-1/3}$ depends on the renormalization-scheme independent terms only of the NLL BFKL kernel with running coupling.
By contrast, the constant term $c_{0}$ \eqref{coeff_c0} is renormalization-scheme dependent via $N_0$, but in such a way that this compensates renormalization-scheme dependence of the $\Lambda_{QCD}$ scale included in the definition of $L_s(Y)$ \eqref{log_Qs}. Finally, one concludes that the asymptotic expression of the saturation scale $Q_s(Y)$ deduced from \eqref{result_Ls_RC} is renormalization-scheme independent.\\

Presumably, 
phenomenological studies or numerical simulations of gluon saturation at NLL accuracy would require a resummation of collinear higher order contributions analog to the one done in the BFKL case \cite{Salam:1998tj,Ciafaloni:1998iv,Ciafaloni:1999yw,Altarelli:1999vw,Altarelli:2001ji,Ciafaloni:2003rd,Altarelli:2005ni}. However, such resummations are formally NNLL effects, so that they cannot modify the coefficients  (\ref{coeff_c0},\ref{coeff_cm16},\ref{coeff_cm13}) of the large rapidity expansion \eqref{result_Ls_RC}.


\section{Applications}\label{sec:plots}

Let us now use the formal results of the previous section to discuss the qualitative effects of running coupling prescriptions and other NLL contributions on the asymptotic behavior of the saturation scale. For the purpose of our discussion, let us take the following values for the remaining parameters: $\kappa=0.1$, ${\cal A}=0.8$, and $N_f=5$. By consistency, one should then take $\Lambda_{QCD}^{\overline{MS}}=0.09\, \textrm{GeV}$, in order to have $\alpha_s(M_Z)=0.1184$ \cite{Bethke:2009jm}.
The evolution of $Q_s(Y)$ is often parameterized by
\be
\lambda(Y) =\f{\dd}{\dd Y} L_s(Y)=\f{2}{{Q_s}(Y)}\: \f{\dd}{\dd Y} {Q_s}(Y)\, ,\label{def_lambda}
\ee
since in the fixed coupling case, $\lambda(Y)$ converges fast towards a constant value. Phenomenological fits of HERA data with models featuring a fixed $\lambda$, inspired by fixed coupling gluon saturation, give typically a value of $\lambda$ around $0.3$ or $0.2$, depending on the details of the model \cite{GolecBiernat:1998js,Bartels:2002cj,Kowalski:2003hm,Iancu:2003ge,Kowalski:2006hc,Soyez:2007kg}.\\

\begin{figure}
\begin{center}
\begin{tabular}{cc}
\includegraphics[width=7cm]{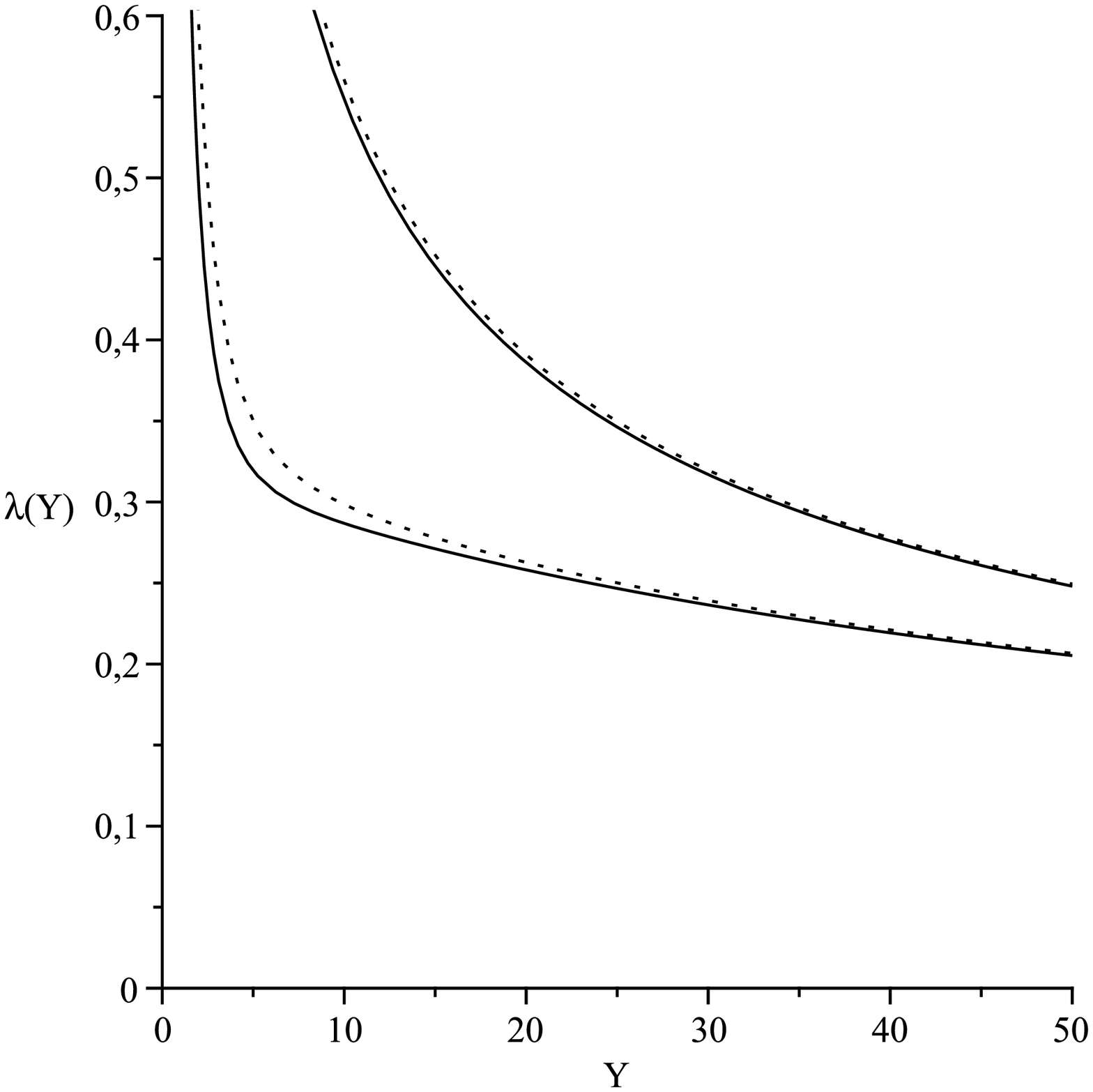} & \includegraphics[width=7cm]{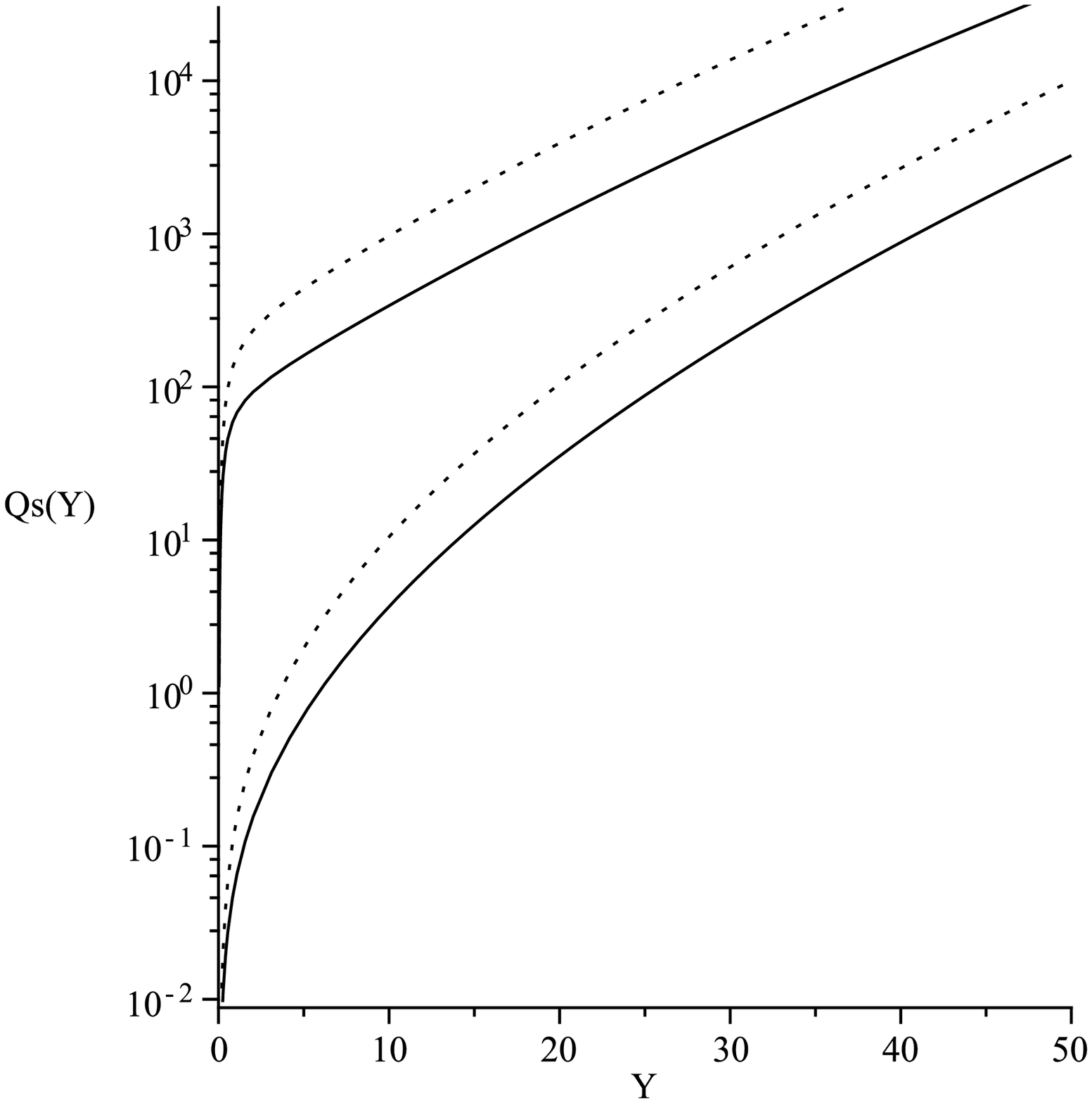}\\
\end{tabular}
\caption{\label{fig:Bal} \emph{Left}: $\lambda(Y)$ (see equation \eqref{def_lambda}) as a function of $Y$. \emph{Right}: The saturation scale $Q_s(Y)$ as a function of $Y$. In both plots, the results for the LL BK equation with running coupling  are drawn with dotted lines for the parent dipole size prescription, \emph{i.e.} with vanishing coefficients $N_n$, and with solid lines for Balitsky's prescription, \emph{i.e.} with the coefficients $N_n$ calculated using the eigenvalue \eqref{chiNLL_Bal}. The upper curves on the left plot (resp. lower curves on the right plot) correspond the expression \eqref{result_Ls_RC} truncated after the term of order $Y^{-1/3}$. By contrast, for the lower curves on the left plot (resp. upper curves on the right plot), a shift $Y\mapsto Y+20$ has been performed in the leading term of the expansion \eqref{result_Ls_RC} only.}
\end{center}
\end{figure}

The results for $\lambda(Y)$ and $Q_s(Y)$ obtained from the expansion \eqref{result_Ls_RC} for improved LL BK equations with running coupling are presented on Fig.\ref{fig:Bal}, both with the parent dipole size prescription, \emph{i.e.} taking $N_0=N_1=N_2=0$,  and with Balitsky's prescription, \emph{i.e.} with the coefficients $N_n$ calculated using the NLL eigenvalue \eqref{chiNLL_Bal}. The rescaling \eqref{LambdaQCDpos} of $\Lambda_{QCD}$ is done here. One can see that the choice of running coupling prescription has only a negligible impact on $\lambda(Y)$. Indeed, the coefficient $c_0$ drops from $\lambda(Y)$ due to the derivation. The three leading terms in the asymptotic expansion of $\lambda(Y)$ are thus independent of the details of the running coupling prescription. Moreover, there are accidental cancelations in the coefficient $c_{-1/3}$ further reducing the sensitivity to the prescription: from the eigenvalue \eqref{chiNLL_Bal}, one has $N_0\simeq -2.45$,  $N_1\simeq -3.03$ and $N_2\simeq -19.4$ but only $N_0\!-\! \g_c N_1\simeq -0.55$ and $N_2\!-\!D N_0\simeq -0.027$. Hence, the main effect of the choice of prescription is a reduction of the value of $Q_s(Y)$ by a factor $\exp(N_0/2)\simeq 0.29$ (or $0.35$ when $N_f=3$) for Balitsky's prescription with respect to the the parent dipole size prescription, as one can see from the right plot of Fig.\ref{fig:Bal}. That seems to be in agreement with numerical simulations (see \emph{e.g.} Fig.4 of Ref.\cite{Albacete:2007yr}).\\

The values of $\lambda(Y)$ obtained with both prescriptions and with no non-universal subleading contributions to the expansion \eqref{result_Ls_RC} are small enough at very large rapidity, as one can see from the left plot of Fig.\ref{fig:Bal}, but remain too large in the phenomenologically interesting range $4<Y<10$. By tuning the non-universal subleading contributions, for example by adding $20$ to $Y$ in the first term of the expansion \eqref{result_Ls_RC}, it is possible to stabilize $\lambda(Y)$ around its phenomenological value down to the range of interest in $Y$. However, such a big shift is not very natural. Moreover, since the asymptotic behavior of $Q_s(Y)$ is universal, reducing the evolution rate $\lambda(Y)$ at intermediate rapidities leads to a significant increase of $Q_s(Y)$ at lower rapidities. As one can see from the right plot of Fig.\ref{fig:Bal}, the value of $Q_s(Y)$ at moderate rapidities is too high\footnote{Notice that our choice $\kappa=0.1$ automatically leads to higher numerical values of $Q_s(Y)$ than the usual definitions of $Q_s(Y)$ in the literature. However, that effect is too small to make the upper curves of the right plot of Fig.\ref{fig:Bal} consistent with the expectations for $Q_s(Y)$.} when the shift is performed. Hence, even by tuning initial condition effects, it seems difficult to get simultaneously the correct values of both $Q_s(Y)$ and $\lambda(Y)$ in the relevant rapidity range from the improved LL BK equation with running coupling, independently of the details of the running coupling prescription. Such difficulty has been bypassed \emph{e.g.} in Ref.\cite{Albacete:2009fh} by multiplying $\Lambda_{QCD}$ by a free parameter fitted on the data.\\

\begin{figure}
\begin{center}
\begin{tabular}{cc}
\includegraphics[width=8cm]{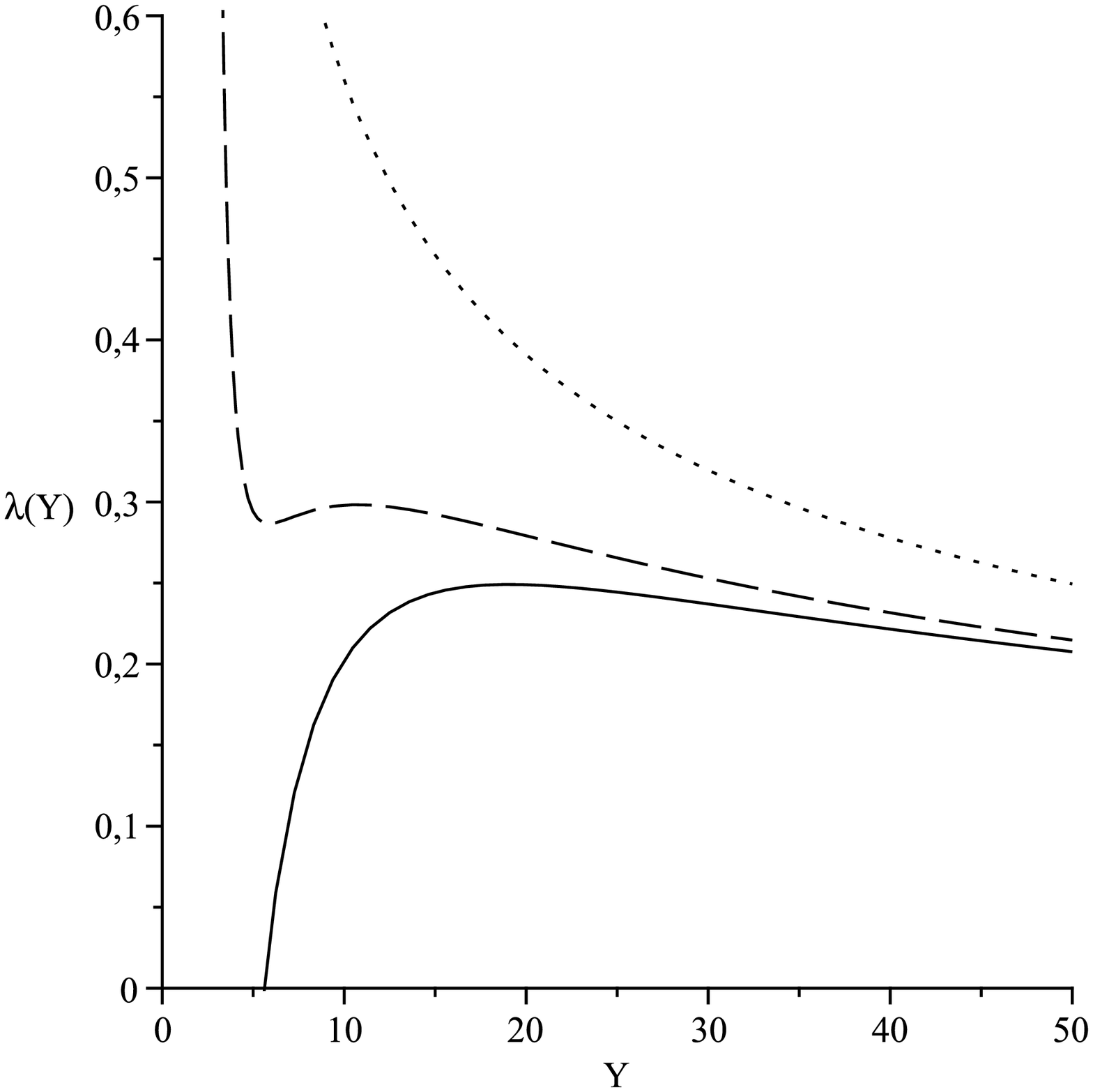} & \includegraphics[width=8cm]{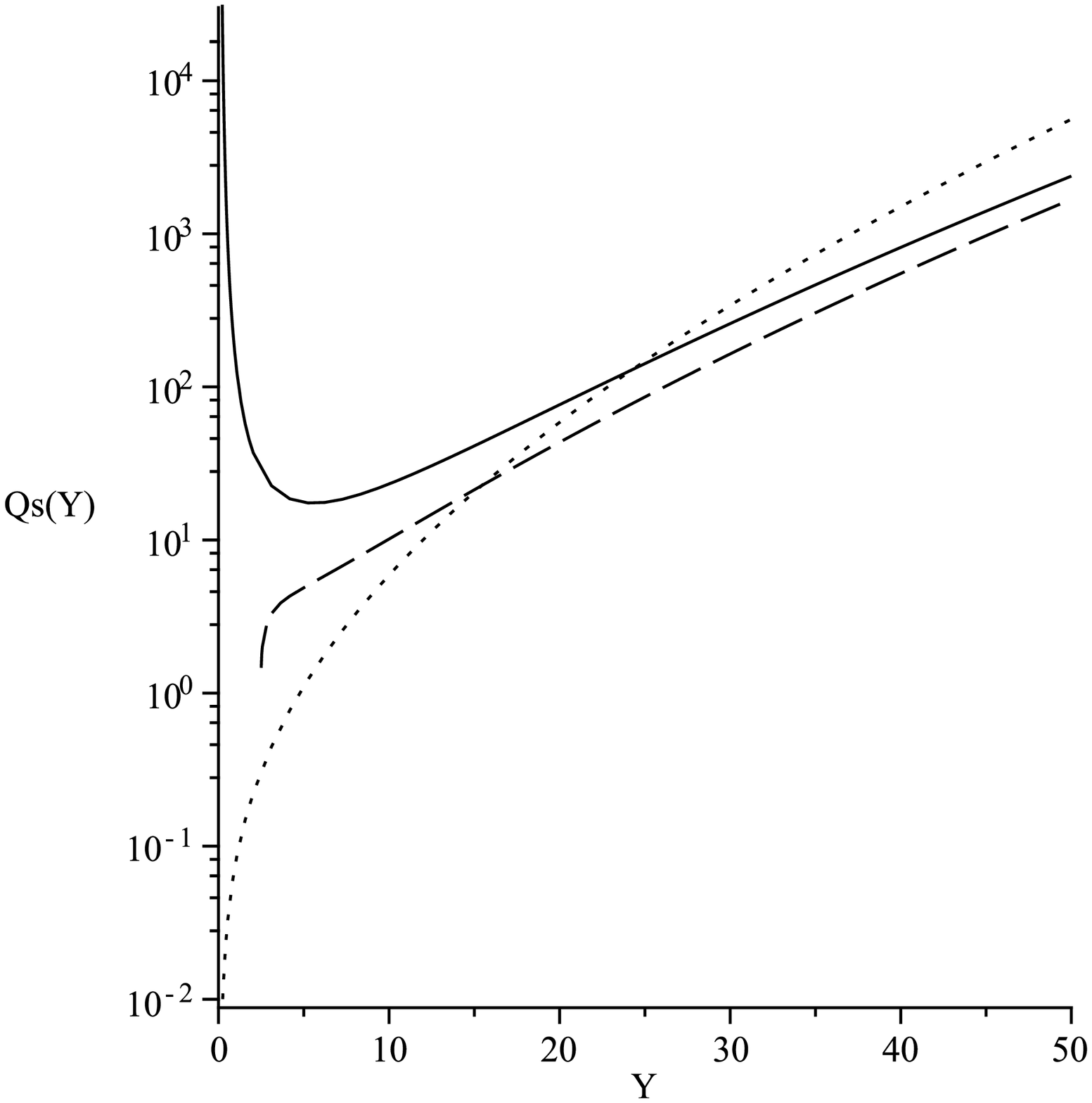}\\
\end{tabular}
\caption{\label{fig:NLL} \emph{Left}: $\lambda(Y)$ (see equation \eqref{def_lambda}) as a function of $Y$. \emph{Right}: The saturation scale $Q_s(Y)$ as a function of $Y$. Both plots are obtained from the expansion \eqref{result_Ls_RC} applied to a unitarized version of the BFKL equation in momentum space with running coupling at the parent gluon $k_\perp$. The full NLL BFKL eigenvalue \eqref{chiNLL_FaLi} is used to calculate the coefficients $N_n$ for the solid and dashed curves, whereas for the dotted curves no NLL contributions are included, \emph{i.e.} the coefficients $N_n$ are taken to be zero. The solid and dotted curves correspond the expression \eqref{result_Ls_RC} truncated after the term of order $Y^{-1/3}$. For the dashed curves, a shift $Y\mapsto Y\!-\!2.5$ has been performed in the leading term of the expansion \eqref{result_Ls_RC} only.
}
\end{center}
\end{figure}

In Fig.\ref{fig:NLL} are presented the results for $\lambda(Y)$ and $Q_s(Y)$ for the BFKL equation in momentum space with running coupling at the parent gluon $k_\perp$, modified by saturation effects, both with the full NLL BFKL eigenvalue \eqref{chiNLL_FaLi} or with the LL one only. One can see on the left hand plot that the NLL corrections on $\lambda(Y)$ are sizable. They stabilize $\lambda(Y)$ down to $Y\simeq 15$, at values around $0.2-0.25$, even without including any non-universal corrections in the expansion \eqref{result_Ls_RC}. In that case, there are indeed no strong cancelation occurring in the NLL contributions to the coefficient $c_{-1/3}$: from the NLL eigenvalue \eqref{chiNLL_FaLi}, one gets $N_0\simeq -7.98$,  $N_1\simeq -3.83$ and $N_2\simeq -128.4$ and thus $N_0\!-\! \g_c N_1\simeq -5.57$ and $N_2\!-\!D N_0\simeq -65.2$. Including moderate nonuniversal subleading contributions in the expansion \eqref{result_Ls_RC} by adding \emph{e.g.} $-2.5$ to $Y$ in the first term further improve NLL results. The values of $\lambda(Y)$ are now compatible with phenomenological results in the relevant rapidity range, where, simultaneously, $Q_s(Y)$ is reduced to small enough values, as we can see on Fig.\ref{fig:NLL}.\\

Let us summarize the results of this section. Compared to the fixed coupling case, running coupling effects significantly improve the agreement between the theoretical predictions and the phenomenological fits for the saturation scale and its evolution. This is expected since the wave front solutions to saturation equations with fixed or running coupling are in different universality classes. However, when including running coupling effects only on top of LL equations, there remains a tension between the theory and the fits, even when strong subleading initial condition effects are included. 
By contrast, when not only the running coupling effects but also the full NLL BFKL effects are included, the theoretical predictions are naturally close to the phenomenological results. In that case, only minor initial condition effects may be required to describe the saturation scale of a proton or of a nucleus, and its rapidity dependance.


\section{Conclusion}

The central result of this paper is the complete universal part of the large rapidity asymptotic expansion of $\log Q_s(Y)$ (\ref{result_Ls_RC},\ref{coeff_c0},\ref{coeff_cm16},\ref{coeff_cm13}), valid for evolution equations with gluon saturation and running coupling. That result is obtained using the traveling wave method of Ref.\cite{vanS}. Among the three new terms in the expansion, two of them are sensitive to the details of the running coupling prescription and other NLL effects, and also to some details of the nonlinear damping via the initial-condition-independent  normalization constant ${\cal A}$ of the universal traveling wave solutions. Initial condition and NNLL effects would appear only at the following order in the expansion of $\log Q_s(Y)$, which is $Y^{-1/2}$, in agreement with previous expectations \cite{Mueller:2003bz,Peschanski:2006bm,Beuf:2007cw}. By contrast to the fixed coupling case, the normalization of the saturation scale $Q_s(Y)$ at large rapidity is thus independent on the nature of the target, and is instead related to $\Lambda_{QCD}$, to the NLL corrections to the BFKL kernel, and to the normalization constant ${\cal A}$ which has to be determined numerically.\\

The details of the running coupling prescription are found to have a negligible impact on the evolution of $Q_s(Y)$, but an important one on its normalization, in agreement with numerical simulations \cite{Albacete:2007yr}. By contrast, the contributions to the NLL BFKL kernel not related to the running of the coupling have a large impact on both the normalization of $Q_s(Y)$ and its evolution, and they allow to obtain results in agreement with phenomenological fits without the need of strong initial condition effects. This remark suggests that the large rapidity expansion \eqref{result_Ls_RC} of $\log Q_s(Y)$ maybe relevant down to the rapidities relevant for collider physics. Further studies are however necessary to fully understand its validity range. The results presented here should also provide a useful constraint on high energy extrapolations of models of hadronic collisions
for ultra high energy cosmic rays.


\begin{acknowledgments}

I would like to thank Robi Peschanski and Raju Venugopalan for useful comments on the manuscript.
This manuscript has been authored under the Contract No. \#DE-AC02-98CH10886 with the
U.S. Department of Energy.

\end{acknowledgments}


\appendix


\bibliography{MaBiblioTW}

\end{document}